\documentclass[a4paper]{jpconf}

\usepackage[english]{babel}

\usepackage[a4paper,top=4cm,bottom=2.7cm,left=2.5cm,right=2.5cm,marginparwidth=1.75cm]{geometry}

\usepackage{amsmath}
\usepackage{graphicx}
\usepackage[colorlinks=true, allcolors=blue]{hyperref}
\begin{document}
\title{$^{10}$Be in the Cluster Shell Model}
\author{Omar Alejandro D\'iaz Caballero}
\address{Instituto de Ciencias Nucleares, Universidad Nacional Aut\'onoma de M\'exico,\\
Coyocac\'an, 04510 Ciudad de M\'exico, Mexico}
\ead{odiazcab@gmail.com}
\author{Roelof Bijker}
\address{Instituto de Ciencias Nucleares, Universidad Nacional Aut\'onoma de M\'exico,\\
Coyocac\'an, 04510 Ciudad de M\'exico, Mexico}
\ead{bijker@nucleares.unam.mx}

%\maketitle

\begin{abstract}
The Cluster Shell Model (CSM)describes light nuclei in terms of $k-\alpha$ particles and $x$ extra nucleons, in which the extra nucleon move in the deformed field generated by the geometric configuration of $\alpha$-particles. We present the first study for the case $x=2$ nucleons in application to  $^{10}Be$ as a cluster of two $\alpha$-particles and two neutrons.
\end{abstract}
\vspace{10mm}
\section{Clusters and the light Nuclei}

The study of the light nuclei goes all the way back to the pioneering contributions of  Wheeler\cite{1} and Hafstad and Teller \cite{2} in the 1930s and  Brink in the 1960s \cite{3}. A recent review can be found in \cite{4}\\ 
In the year 2000, an algebraic method called the Algebraic Cluster Model (ACM), was proposed to describe light nuclei as clusters of $\alpha$-particles \cite{5}. As an example, the nucleus $^{12}C$ was described successfully in the ACM as a triangular configuration of three $\alpha$-particles \cite{6}, in particular, the $L^P=5^-$ state was predicted more than a decade before its experimental discovery \cite{7}.\\
In subsequent years, the Cluster Shell Model (CSM) was introduced to describe the properties of neighboring cluster nuclei of the type $k\alpha+x$ as composed of $k $ $\alpha$-particles plus $x=1$ extra nucleons \cite{8,9,10,11}. This contribution aims to present the first application for $x=2$ extra nucleons. As an example, we study the nucleus $^{10}Be$ as a cluster of $k=2$ $\alpha$-particles and $x=2$ neutrons and investigate what  extent  the cluster structure of two $alpha$-particles persist under the addition of two neutrons

\section{Cluster Shell Model}
 
 The Cluster Shell Model describes configurations of light nuclei of the type $ k\alpha + x$ i.e. composed of $k$ $\alpha$-particles plus $x$ extra nucleons, in which the extra nucleons move in the deformed field generated by the cluster of $\alpha$-particles \cite{8}. The case of one extra nucleon with $x=1$ was studied for $^9Be$ and $^9B$ ($k=2$) \cite{9}, $^{13}C$ ($k=3$) \cite{10}, and $^{21Ne}$ and $^{21}Na$ ($k=5$) \cite{11}.\\
In this contribution, we present the first application for two extra nucleons for the case of the nucleus $^{10}Be$ as a cluster of $k=2$ $\alpha$-particles and $x=2$ neutrons. The neutron single-particle levels are described by the CSM Hamiltonian 
\begin{equation}
H=T +   V(\vec{r})+V_{so,r}+  \frac{1}{2}(1+\tau_3) V_C(\vec{r})
\end{equation}
which is the sum of the kinetic energy, a central potential, a spin-orbit interaction, and a Coulomb potential in the case of an odd proton. The central potential is obtained by convoluting the density
\begin{equation}
 \rho(\vec{r})= \sum_{i=1}^k\Big(\frac{\alpha_i}{\pi}\Big)^\frac{3}{2}e^{-\alpha_i (\vec{r}-\vec{r_i})^2}
    \end{equation}
with the nucleon-$\alpha$ interaction \cite{7}, The coefficient $\alpha$ is related to the size of the $\alpha$-particle, and the $\vec{r}_i$ represents the coordinates of the $\alpha$-particles: $\Vec{r}_1=(\beta,0,-)$ and $\Vec{r}_2=(\beta,\pi,-)$ where $\beta$ represents the distance of the $\alpha$ particles with respect to the center of mass.\\
 The CSM wave functions are calculated in the intrinsic or body-fixed frame. A system of two identical $alpha$-particles has axial symmetry, and hence the eigenstates are characterized by  $K^P$: the projection of the angular momentum along the symmetry axis $K$ and the parity $P$. The calculations are carried out on the harmonic oscillator basis. The $i$-th eigenstate is given by the expansion:
 \begin{equation}
\mid  K^P_i \rangle=\sum_{n,l,j}c_{nljm}^{K(i)}\mid n,(l,\frac{1}{2}),j,K\rangle\delta_{P,(-1)^l}
\end{equation}
Figure 1 shows the splitting of the spherical single-particle levels in the deformed field generated by the clusters of two $\alpha$-particles as a function of $\beta$ \cite{8,9}. For $\beta=0$ we recover the spherical harmonic limit, whereas for $\beta>0$ the harmonic oscillator levels are mixed and split into the levels characterized by $K^P$. For the case of interest, the value of $\beta$ is determined from the moment of inertia of the rotational band in $^8Be$ as $\beta =1.82 $fm \cite{9}. The five lowest eigenstates are listed in table  \ref{fig:table1}.
\begin{table}[h!]
\caption{\label{EE}Eigenvalues of CSM in MeV}
\begin{center}
\begin{tabular}{c c c}
\hline
 & $K^P$ & Energy \\\hline
$1s_{\frac{1}{2}}$  & $ \frac{1}{2}^+$ &-10.38 \\
$1p_{\frac{3}{2}}$ & $\frac{1}{2}^-$ &-3.81\\
$1p_{\frac{3}{2}}$ & $\frac{3}{2}^-$ &-1.75\\
$1p_{\frac{1}{2}}$ &  $\frac{1}{2}^-$ &0.35 \\
$1d_{\frac{5}{2}}$ & $\frac{1}{2}^+$ &1.38\\
\hline
\end{tabular}
\end{center}
\label{fig:table1}
\end{table}
 \begin{figure}[h!]
  \centering
     \includegraphics[width=0.55\linewidth, height=0.45\textheight]{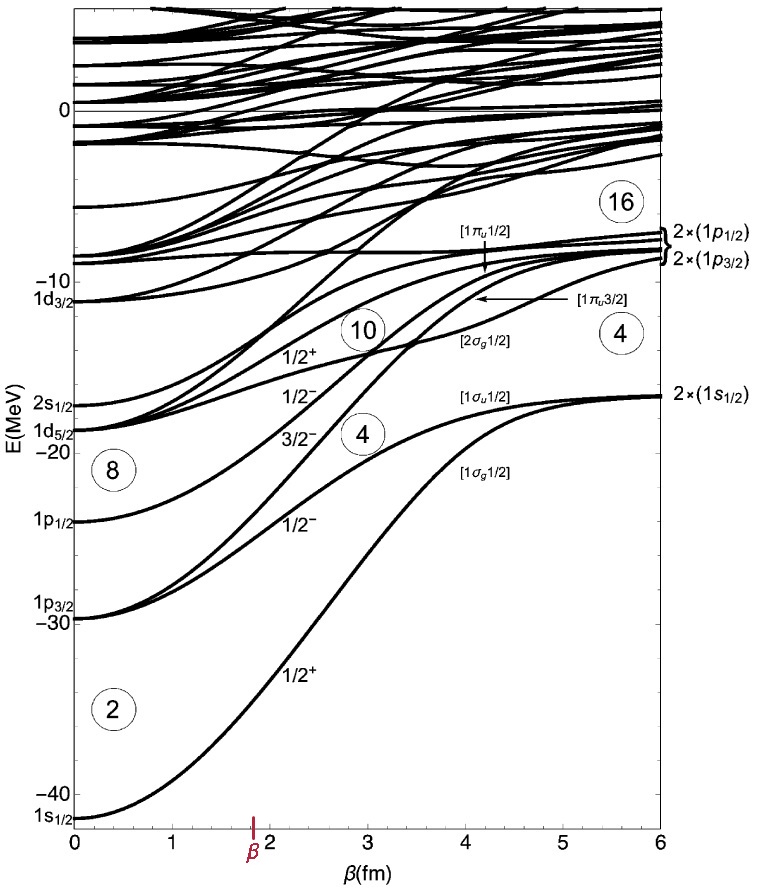}
      \caption{Excited states of the $^9Be$ as a function of  $\beta$}
     \label{fig:exitedBe}
 \end{figure}  
 \section{The nucleus $^{10}Be$}
In this section, we discuss the nucleus $^{10}Be$ as a cluster of $k=2$ $\alpha$-particles and $x=2$ neutrons. In addition to the CSM Hamiltonian of Eq. (1) we have to add the residual interaction between the two neutrons. Here we propose a pairing interaction \cite{12}
  \begin{eqnarray}
V_{pair}=-G\sum_{\mu,\nu>0}a^\dagger_\mu a^\dagger_{\bar{\mu}} a_\nu a_{\bar{\nu}}\nonumber\\
=-G\sum_{\mu,\nu>0}Q_+(\mu)Q_-(\nu)
\end{eqnarray}

In this notation $\bar{\mu}$ represents the time-reversed partner of $\mu$ which is degenerate with $\mu$. The pair creation and annihilation operators,$Q_+(\mu)$ and $Q_-(\nu)$, together with the number operator  :
\begin{equation}
Q_0(\mu)=\frac{a^\dagger_\mu a_\mu-a_{\bar{\mu}}a^\dagger_{\bar{\mu}}}{2}
\end{equation}
satisfy the $SU(2)$ fermion quasi-spin algebra
\begin{eqnarray}
 [Q_+(\mu),Q_-(\nu)]=2\delta_a^b Q_0(\mu)\nonumber\\
 \left[  Q_0(\mu),Q_\pm(\nu) \right] =\pm\delta_a^b Q_\pm(\mu)
 \end{eqnarray}
The two-particle states are given by:
\begin{equation}
\mid{\mu}\rangle=Q_+(\mu)\mid 0 \rangle \quad \langle \mu \mid = \langle 0 \mid Q_-(\mu)
\end{equation}
The matrix elements of the pairing interaction in the two-particle states are:
\begin{eqnarray}
    \langle V_{pair}\rangle =  -G \sum_{\mu',\nu'>0} \langle 0 \mid Q_- (\mu) Q_+(\mu')Q_-(\nu)Q_+(\nu')\mid 0 \rangle\nonumber
  \end{eqnarray}
  \begin{equation}
      =-G
  \end{equation}
The relevant single-particle levels are shown in Table \ref{fig:table1}. The first two levels with $K^P=\frac{1}{2}_1^+$ and $\frac{1}{2}_1^-$ are occupied by the four neutrons of the two $\alpha$-particles. In our calculation, we consider the next three available levels with  $K^P=\frac{3}{2}_1^-$,  $\frac{1}{2}_2^-$ and  $\frac{1}{2}_2^+$ for the two extra neutrons.

\begin{figure}[t]
\includegraphics[width=0.75\linewidth, height=0.3\textheight]{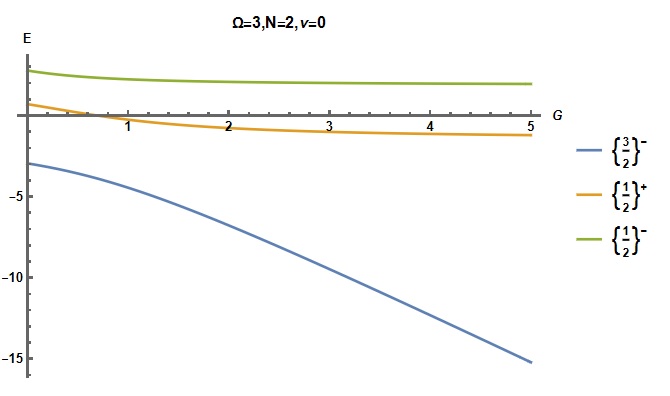}
\caption{Eigenvalues corresponding of $J=0^+
$ as a function of the intensity of the pairing strength G}
\label{fig:PI}
\end{figure} 

\begin{figure}[b]
\includegraphics[width=0.8\linewidth, height=0.35\textheight]{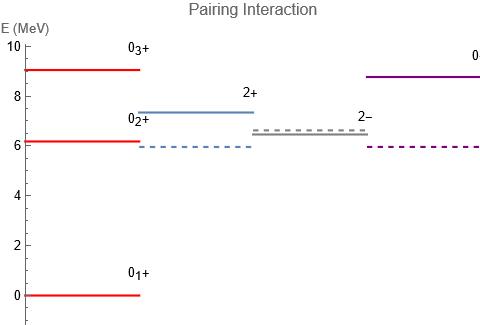}
\caption{Eigenvalues given by the Paring interaction and the experimental data (dotted line)}
\label{fig:PairData}
\end{figure}     

The only configurations affected by the pairing interaction are the $K^P=0^+$ states. The eigenvalues can be obtained by diagonalizing the CSM Hamiltonian plus the pairing interaction, $H_{CSM}+V_{pair}$, with matrix elements.
The matrix elements of the pairing interaction are then:
\begin{equation}
\begin{pmatrix}
-G+2E_3 & -G & -G \\
-G & -G+2E_4 & -G \\
-G & -G & -G+2E_5 
\end{pmatrix}
\end{equation}
Figure \ref{fig:PI} shows the behavior of the $K^P=0^+$ eigenstates as a function of the pairing strength $G$. The value of the pairing strength is determined from the excitation energy of the first exited $0^+$ state to be $G=2.079$Mev the final results for $^{10}Be$ are shown in Figure \ref{fig:PairData}.

\section{Summary and conclusions}
We presented the first extension of the Cluster Shell Model to a cluster of two $\alpha$-particles with two extra nucleons with an application to the nucleus $^{10}Be$. The residual interaction between the two extra neutrons was taken as a pairing potential whose strength was adjusted to the excitation energy of the first $0^+$ state. We find a good agreement with the available experimental data. A more detailed CSM analysis of $^{10}Be$ including other forms of residual interaction as well as a study of transition probabilities will be published elsewhere.  
\section*{Acknowleddgments}
This work was supported in part by grant No. 1G101423 from PAPIIT, DGAPA-UNAM, Mexico.
\section*{References}
\bibliographystyle{alpha}

\begin{thebibliography}{9}
\bibitem{1}
J.A. Wheeler, Phys. Rev. 52 (1937) 1083.
\bibitem{2}
L.R. Hafstad, E. Teller, Phys. Rev. 54 (1938) 681.
\bibitem{3}
D.M. Brink, in: Proc. Int. School of Physics “Enrico Fermi”, Course XXXVI, 1965, p.247.
\bibitem{4}
M. Freer, H. Horiuchi, Y. Kanada-En’yo, D. Lee, and U-G. Meissner, Rev. Mod. Phys. 90, (2018) 035004.
\bibitem{5}
R.Bijker and F. Iachello Physical Review C 61(2000) 067305
\bibitem{6}
Bijker R and Iachello F, 2002 \textit{Annals of Physics} 298, 334-360 
\bibitem{7}
D. J. Marín-Lámbarri,  Bijker R., M. Freer, M. Gai, Tz. Kokalova, D.J Parker and C. Wheldon, Physical Review Letters, 113 (2014) 012502
\bibitem{8}
V.Della Rocca, R.Bijker,  and F.Iachello (2017) \textit{ Nuclear PhysicsA}   966 158–184
 \bibitem{9}
V.Della Rocca and F.Iachello (2018) \textit{ Nuclear PhysicsA}   973 1–32
\bibitem{10}
R. Bijker and F. Iachello, Physical Review Letters, 122 (2019) 162501
\bibitem{11}
R. Bijker and F. Iachello, Nuclear Physics A. 1010(2021)122193
\bibitem{12} 
O. Burglin, N. Rowley, Nuclear Physics A. 602(1996) 21-40
\bibitem{13}
O. A.  Díaz, and R. Bijker in preparation
\end{thebibliography}

\end{document}